\documentstyle[epsfig]{mn}

\newif\ifAMStwofonts

\ifoldfss
  \newcommand{\rmn}[1] {{\rm #1}}

  \ifCUPmtlplainloaded \else
    \NewTextAlphabet{textbfit} {cmbxti10} {}
    \NewTextAlphabet{textbfss} {cmssbx10} {}
    \NewMathAlphabet{mathbfit} {cmbxti10} {} 
    \NewMathAlphabet{mathbfss} {cmssbx10} {} 
  \fi
  \ifAMStwofonts
    \ifCUPmtlplainloaded \else
      \NewSymbolFont{upmath} {eurm10}
      \NewSymbolFont{AMSa} {msam10}
      \NewMathSymbol{\upi}     {0}{upmath}{19}
      \NewMathSymbol{\umu}     {0}{upmath}{16}
      \NewMathSymbol{\upartial}{0}{upmath}{40}
      \NewMathSymbol{\leqslant}{3}{AMSa}{36}
      \NewMathSymbol{\geqslant}{3}{AMSa}{3E}

       \let\le=\leqslant
       
    \fi
  \fi
\fi 

\ifnfssone
  \newmathalphabet{\mathit}
  \addtoversion{normal}{\mathit}{cmr}{m}{it}
  \addtoversion{bold}{\mathit}{cmr}{bx}{it}
  \newcommand{\rmn}[1] {\mathrm{#1}}

  \newmathalphabet{\mathbfit} 
  \addtoversion{normal}{\mathbfit}{cmr}{bx}{it}
  \addtoversion{bold}{\mathbfit}{cmr}{bx}{it}
  \newmathalphabet{\mathbfss} 
  \addtoversion{normal}{\mathbfss}{cmss}{bx}{n}
  \addtoversion{bold}{\mathbfss}{cmss}{bx}{n}
  \ifAMStwofonts
    \ifCUPmtlplainloaded \else
      \UseAMStwoboldmath
      \makeatletter
      \new@mathgroup\upmath@group
      \define@mathgroup\mv@normal\upmath@group{eur}{m}{n}
      \define@mathgroup\mv@bold\upmath@group{eur}{b}{n}
      \edef\UPM{\hexnumber\upmath@group}
      \new@mathgroup\amsa@group
      \define@mathgroup\mv@normal\amsa@group{msa}{m}{n}
      \define@mathgroup\mv@bold\amsa@group{msa}{m}{n}
      \edef\AMSa{\hexnumber\amsa@group}
      \makeatother
      \mathchardef\upi="0\UPM19
      \mathchardef\umu="0\UPM16
      \mathchardef\upartial="0\UPM40
      \mathchardef\leqslant="3\AMSa36
      \mathchardef\geqslant="3\AMSa3E

       \let\le=\leqslant

    \fi
  \fi
\fi 

\ifnfsstwo
  \newcommand{\rmn}[1] {\mathrm{#1}}

  \DeclareMathAlphabet{\mathbfit}{OT1}{cmr}{bx}{it}
  \SetMathAlphabet\mathbfit{bold}{OT1}{cmr}{bx}{it}
  \DeclareMathAlphabet{\mathbfss}{OT1}{cmss}{bx}{n}
  \SetMathAlphabet\mathbfss{bold}{OT1}{cmss}{bx}{n}
  \ifAMStwofonts
    \ifCUPmtlplainloaded \else
      \DeclareSymbolFont{UPM}{U}{eur}{m}{n}
      \SetSymbolFont{UPM}{bold}{U}{eur}{b}{n}
      \DeclareSymbolFont{AMSa}{U}{msa}{m}{n}
      \DeclareMathSymbol{\upi}{0}{UPM}{"19}
      \DeclareMathSymbol{\umu}{0}{UPM}{"16}
      \DeclareMathSymbol{\upartial}{0}{UPM}{"40}
      \DeclareMathSymbol{\leqslant}{3}{AMSa}{"36}
      \DeclareMathSymbol{\geqslant}{3}{AMSa}{"3E}

       \let\le=\leqslant

    \fi
  \fi
\fi 

\ifCUPmtlplainloaded \else
  \ifAMStwofonts \else 
    \def\upi{\pi}
    \def\umu{\mu}
    \def\upartial{\partial}
  \fi
\fi

\title{The nature of the dwarf population in Abell~868}
\author[P.J. Boyce et al.]
       {Peter J. Boyce,$^1$ Steven Phillipps$^1$, J. Bryn Jones$^2$, Simon P. Driver$^3$ 
 \newauthor Rodney M. Smith$^4$ and Warrick J. Couch$^5$ 
      \\ 
   $^1$Astrophysics Group, Department of Physics, University of Bristol, Tyndall Avenue, Bristol, BS8 1TL\\
   $^2$Astronomy Group, School of Physics and Astronomy, University of Nottingham,
 University Park, Nottingham, NG7 2RD\\
   $^3$School of Physics and Astronomy, University of St Andrews, North Haugh, St Andrews, KY16 9SS\\ 
   $^4$Department of Physics and Astronomy, University of Wales Cardiff, P.O. Box 913, Cardiff, CF2 3YB \\
 $^5$School of Physics, University of New South Wales, Sydney, NSW 2052, Australia}

\date{Accepted ???.
      Received ???? ;}

\pagerange{\pageref{firstpage}--\pageref{lastpage}}
\pubyear{1994}

\begin{document}

\maketitle

\label{firstpage}

\begin{abstract}
We present the results of a study of the morphology of the 
 dwarf galaxy population in Abell 868, a rich, intermediate 
redshift ($z$=0.154) cluster which has a galaxy luminosity function with a 
 steep faint-end slope ($\alpha$=--1.26$\pm$0.05). 
  A statistical background subtraction 
 method is employed to study the $B-R$ colour distribution of 
  the cluster galaxies.  
 This distribution suggests 
 that the galaxies contributing to the faint-end of the measured 
 cluster LF 
  can be split into three populations: dIrrs with $B-R<$1.4; dEs with 
  1.4$\le$$B-R$$\le$2.5; and contaminating background giant ellipticals 
 (gEs) with $B-R>$2.5. The removal 
 of the contribution of the background gEs from the counts
 only marginally lessens the faint-end slope ($\alpha$=--1.22$\pm$0.16). 
 However, the removal of the contribution of the  dIrrs from the counts 
   produces a flat LF ($\alpha$=--0.91$\pm$0.16). 
  The dEs and the dIrrs have 
 similar spatial distributions within the cluster except that the 
 dIrrs appear to be totally absent within a central projected radius 
 of about 0.2~Mpc ($H_{o}$=75~km$\,$s$^{-1}$~Mpc$^{-1}$). The 
   number densities of both dEs and dIrrs 
   appear to fall off beyond a projected radius of $\simeq$0.35~Mpc. 
  We suggest 
  that the dE and dIrr populations of A868 have been associated with 
 the cluster for similar timescales but that evolutionary processes such as 
 `galaxy harassment'  tend to fade the dIrr galaxies while  
 having much less effect on the dE galaxies. The harassment would be expected to 
  have the greatest effect on dwarfs residing in the central parts of the cluster. 
\end{abstract}

\begin{keywords}
galaxies: clusters: general -- galaxies: evolution -- galaxies: dwarf -- galaxies: luminosity function, mass function -- methods: data analysis.
\end{keywords}

\section{Introduction}

The galaxy luminosity function (LF)  is one of the most direct 
 observational tests of theories of 
 galaxy formation and evolution. Clusters of galaxies are ideal
  systems within which to measure the 
 galaxy LF down to very faint magnitudes because of the large numbers of galaxies at the same 
 distance that can be observed within a small area of sky. 
 Much work has been done in recent years in measuring the faint-end of the galaxy LF 
 in clusters (e.g. Driver et al. 1994; de Propris et al. 1995;  
 Lobo et al. 1997; Wilson et al. 1997; Valotto et al. 1997; 
  Smith, Driver \& Phillipps 1997; Trentham 1997a, 1997b, 1998; de Propris \& Pritchett 1998; Driver, Couch \&
 Phillipps 1998b; Garilli, Maccagni \& Andreon 1999). These studies 
 have generally used background subtraction procedures 
  and have shown that   many clusters 
   are dominated in number by a large population of faint 
 galaxies. The LF of these clusters typicially becomes steep 
 faintward of about  M$_{R}$$\simeq$--18.1.
  The faint-end slope of the LF, $\alpha$ (where 
 $\alpha$ is the slope of $dN/dL$: $\alpha$=--1 is flat in a plot of 
 log N vs. magnitude) 
 typically lies  in 
the range --1.2 to --2.2.

 Phillipps et al. (1998) noted that the steepness of the 
  faint-end slope appears to be dependent on cluster 
 density, with dwarfs being more common in lower density environments. 
 This is possibly because the various dynamical processes which can 
 destroy dwarf galaxies act preferentially in dense environments.  
 One implication of this is that  dwarfs may 
 be less common in the higher density cores of clusters than at larger 
 cluster radii. For example, Lobo et al. (1997) measured 
 $\alpha$=--1.8 from a large area (1500~arcmin$^2$) survey of the Coma cluster. 
 However,  Adami et al. (2000) 
  derived a flat faint-end LF from a spectroscopic 
 survey of a small area
  (56~arcmin$^2$) around the core of the cluster. Driver et al. 
  (1998b) noted a tendency for less evolved clusters 
 (including A868) to have steeper faint-end slopes compared to more evolved 
 clusters (see also Lopez-Cruz et al. 1997).  The implication is that 
  a larger fraction of dwarfs has been destroyed in the more evolved clusters. 
  
   Measurements of the
  field galaxy LF from redshift surveys 
 (e.g. Loveday et al. 1992; Lin et al. 1997; 
 Marzke et al. 1997; Bromley et al. 1998; 
 Muriel et al. 1998)  generally  give a flat faint-end slope with 
 $\alpha\simeq$--1.0$\pm$0.1. However, other recent measurements of the 
 field galaxy LF have found a steeper LF $\alpha$$\simeq$--1.2 (Zucca et al 1997; 
 Folkes et al. 1999). 
  The LF of the Local Group is flat 
 to $\simeq$L$_{\star}$/10000 (van den Bergh 1992; Pritchet \& van den Bergh 1999). 
 Results from nearby diffuse groups (e.g.  Trentham, Tully \& Verheijen 2001) 
 also give relatively flat faint-end slopes to the LF.

 Hierarchical clustering theories of galaxy formation generically predict a steep mass function 
 of galactic halos (e.g. Kauffmann, White \& Guideroni 1993; Cole et al. 1994). 
 This is in conflict with the flat galaxy LF measured in the field and in 
 diffuse local groups but not with the steep LFs measured in many 
 clusters. However, in the hierarchical universe, clusters form relatively recently 
 from the accretion of smaller systems. 
  The dynamical processes that operate 
 in clusters are destructive. Ram pressure stripping (e.g.
  Abadi, Moore \& Bower 1999) and gravitational tides/galaxy harassment (e.g. 
 Moore et al. 1996; Moore, Lake \& Katz 1998; Bekki, Couch \& Shioya 2001) 
   will both tend to fade galaxies by removing 
 gas or stripping stars. These processes are most effective for less massive, 
  less 
 bound systems. Hence, we might expect to see a
  flattening of the faint-end slope
 in clusters compared to the field, rather than the observed steepening.

Two selection effects may go some way to resolving this apparent paradox. Firstly, 
 surface brightness selection effects may be seriously affecting the derived
 LFs both in clusters and the field (see e.g. Impey \& Bothun 
 1997; Cross et al. 2001). Secondly, 
  Valotto, Moore \& Lambas (2001) have used a numerical simulation of a 
 hierarchical universe to show that  
  many ``clusters'' identified from 
    two dimensional galaxy distributions may result principally 
   from the projection 
   of large-scale structure along the line of sight. 
 They suggest that attempts 
 to derive a LF for these ``clusters'' using the standard background subtraction 
 procedure lead to a derived LF with a steep faint-end slope, despite the fact that the
 actual input LF  had a flat faint-end. Further work needs to be 
 done to establish whether, and to what extent, these effects are biasing measured  
  LFs. 

 Assuming that selection effects cannot completely explain the 
 dichotomy between the steep LFs seen in clusters and the flat LFs 
 seen in the field and loose groups, then it is vital that the nature of the galaxies 
 causing the steep faint-end slope in clusters be studied in detail. 
 The origin and evolution of these objects may have important consequences for our 
 understanding of galaxy formation and the processes which drive galaxy evolution. 

 In this paper, we present the results of a study of the colour characteristics
  of the 
  dwarf population of the cluster A868. 
   A868 is an Abell richness class 3 cluster (Abell, Corwin \& 
 Olowin 1989) at redshift $z$=0.154 (Kristian, Sandage \& Westphal 1978). 
 Its  Bautz-Morgan class is II-III (Leir \& van den Bergh 1977).  
 Driver et al. (1998b) 
 measured a steep faint-end slope to the (R-band) LF for this cluster. 
  The presence 
 of a giant cD galaxy (Valentijn \& Bijleveld 1983)  
   and  the richness of this cluster   suggest that 
  it is 
 not simply the result of the projection of large-scale structure along 
 the line of sight.

 In this paper we use the $B-R$ colour distribution of the cluster galaxies 
 to derive some broad morphological information about the 
 dwarf population of the cluster. In Section 2 
 we discuss the observations and data reduction methods employed. In Section 3
 we present the derived $B-R$ colour distribution of the galaxies at the faint-end
 of the cluster LF. This enables us to separate out the dE and dIrr populations 
   and study their relative contributions to the cluster LF (Section 4). 
   In Section 4 we also discuss the radial distribution of the dE and dIrr populations
 within the cluster.

Throughout this paper we have adopted a standard flat cosmology 
 with $\Omega_{o}$=1, $q_{o}$=0.5, 
and  $H_{o}$=75~km$\,$s$^{-1}$~Mpc$^{-1}$. 

\section{Observations and Data Analysis}

\subsection{Observations and data reduction}

Deep B and R images were obtained of a field centered on 
 the cluster A868 and of a `background' field
  74~arcmin east of the cluster centre. A standard Kitt Peak 
  filter set was used.  The data were obtained at 
  the f/3.3 prime focus of the 3.9m Anglo-Australian 
 Telescope (AAT) using
  a 1024$\times$1024 24-$\mu$m (0.38-arcsec) pixel thinned Textronix CCD. 
 This detector gave a total field of 
 view of $\simeq$6.7$\times$6.7~arcmin$^{2}$ (equivalent to a projected 
 size of $\sim$0.9~Mpc$\times$0.9~Mpc at the distance of A868). 
  The R-band data were obtained on 1996 January 25.  The B-band
  data were obtained  on 1996 
  February 15 and 17 with exactly the same observing set-up.

 The data were reduced using the Starlink CCDPACK  package to de-bias, flat-field, 
 align and co-add the images. 
 The reduced frames were then sky-subtracted 
 to remove any  large-scale residual sky gradients. 
  The R-band data were calibrated using observations of standard star fields from 
 the lists of Landolt (1992) putting our photometry on the 
 Johnson (B)-Cousins (R) magnitude   
  system. The total exposure times for the final co-added 
 R-band frames for both cluster and background fields is 90~min. 
   Neither of 
 the nights on which the B data were taken was photometric. These
   data were calibrated using 
  photometric data obtained on the Jakobus Kapteyn Telscope on La Palma in 
 2000 November. The total exposure time for the final co-added B-band cluster and
  background frames are 240~mins and 180~mins respectively.

\begin{figure}
  \epsfig{file=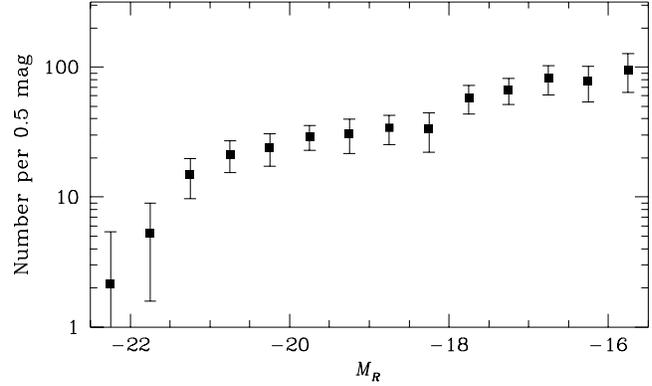}
  \caption{The derived  R-band LF for Abell 868.}
\end{figure}

\subsection{Image detection and photometry}

The detection and photometry of objects on the final co-added 
 R-band cluster and background frames  
  were conducted automatically using the SExtractor software package 
 (Bertin \& Arnouts 1996). A detection limit of $\mu_{\rmn R}$=26~mag~arcsec$^{-2}$ 
 over 4 connected pixels was used.
 The magnitudes measured were Kron (1978) types taken within an aperture of 
 radius = 3.5$r_{kron}$. The measured magnitudes were corrected for 
 Galactic extinction using values from the NED Galactic extinction calculator based upon 
  the maps of Schlegel, Finkbeiner \& Davis (1998).  These values are 
  $A_{R}$=0.09, $A_{B}$=0.22  for the cluster frames and  $A_{R}$=0.15, 
  $A_{B}$=0.36 for the background frames.  
  The positions and apertures defined by SExtractor's   
  source detection algorithm as run on the R-band frame were then used 
  on the B frames. In this way we measured  
  the apparent $B$ magnitudes for those objects detected on the R frames inside 
 identical apertures.

\subsection{Number counts and the luminosity function}

 Driver et al. (1998b) showed, by comparing the counts from 
 the R-band background data to those of Metcalfe 
 et al. (1995), that these data are complete to at least $R$=23.5
 (to Metcalfe et al.'s surface brightness limit of $\simeq$25.2~R$\mu$). 
 We actually use  data 
   to $R$=23.52 since this is equivalent to $M_{R}$=--15.5 in 
   our adopted cosmology.  
   To this completeness limit we expect the measured apparent $R$ 
 magnitudes to have a random error of less than  $\pm$0.1~mag 
 (Driver et al. 1998a). However, we 
  expect magnitudes brighter than 
 $R$=22.5 to have a random error of less than $\pm$0.05~mag.

All of the objects with $R<$23.52 have a measured apparent $B$ magnitude. 
Objects can be detected on the B frames to about 1 magnitude fainter
than on the R frames with similar S/N. Hence, at $R$=22.5 we expect the measured 
 $B-R$ colours to have a random error of around $\pm$0.07~mag if $B-R$=1, 
 $\pm$0.11~mag if $B-R$=2 and $\pm$0.21~mag if $B-R$=3. 
 At $R$=23.5 we expect  the measured 
  $B-R$ colours to have a random error of around $\pm$0.14~mag if $B-R$=1, 
  $\pm$0.23~mag if $B-R$=2 and  $\pm$0.35~mag if $B-R$=3.

Figure 1 presents an R-band LF derived from the data. This was recovered using 
  the standard method of 
 statistically subtracting  the field galaxy counts 
 from those observed towards 
 the cluster (see e.g. Driver et al. 1998b). 
  We have not attempted to apply a k-correction to any 
 of the objects since we do not want to make assumptions about their 
 morphologies before studying the $B-R$ colour distribution. 
  We did make an adjustment for 
 the diminishing field of view available to fainter objects 
 (see Smith et al. 1997). Aside from poisson statistics, 
 we have estimated a further uncertainty of 15\% in the 
 number of background counts, to account for  field-to-field 
  variance due to large-scale structure. This estimate is based 
 upon  Driver et al.'s (1998b) R-band counts for 7 
 non-cluster fields
  made during the same run as the A868 R-band images were 
 taken (see their fig.~3). 
  The derived LF shows
   a sharp upturn faintward of $M_{R}$=--18.0 ($\alpha$=--1.26$\pm$0.05), as 
   noted previously by Driver et al. (1998b). 

\begin{figure}
  \epsfig{file=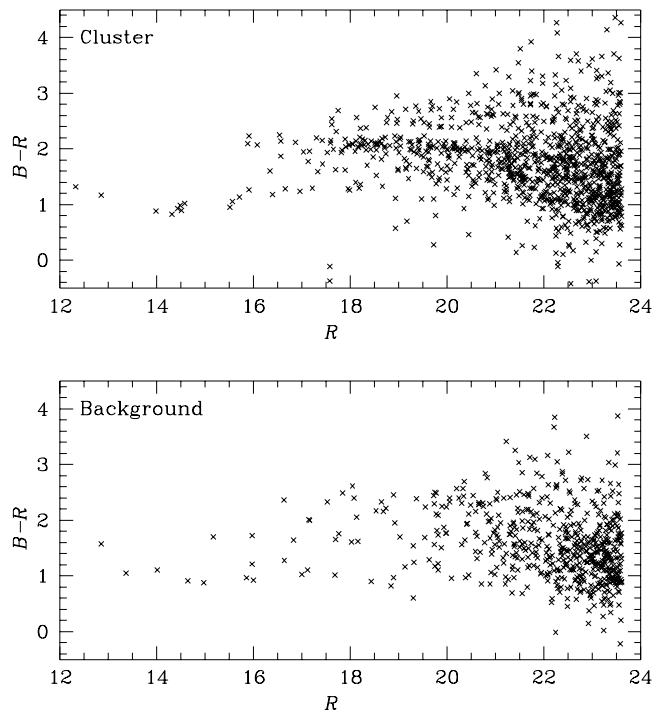}
  \caption{Colour-magnitude diagrams for the cluster and the background frames.}
\end{figure}

Figure~2 shows a colour-magnitude ($R$ versus $B-R$) diagram for 
   the cluster frame and for the background frame.
   The cluster giant elliptical sequence at $B-R\simeq$2.2 
 is clearly visible. This is the observed $B-R$ we expect for these galaxies 
using the k-corrections of Coleman, Wu \& Weedman (1980) and the redshift 
 of A868 ($z$=0.154). 

\section{The $B-R$ Colour distribution}

In this section we consider what the $B-R$ colour data can tell us about the 
 population of galaxies responsible for the upturn seen in the LF at  
  $M_{R}>$--18.0. Figure~3a presents a histogram of the $B-R$ colours for all 
 objects in the cluster frame with 21.02$<R<$23.52 (equivalent to --18.0$<M_{R}<$--15.5), 
  i.e. the apparent magnitude range covering the upturn in the 
 LF.  Figure~3b shows a histogram of the $B-R$ 
 colours for all objects from the same $R$ magnitude range on the background frame. 
 Figure~3c shows the result of subtracting the background frame $B-R$ colour 
 distribution from the cluster frame $B-R$ colour distribution. If one assumes that the 
 population of field galaxies is identical in both frames, not just in 
  its luminosity distribution but also in morphology and hence colour, then the 
 result of the subtraction gives the $B-R$ distribution of the cluster minus 
 the background field population. As with the LF in Fig.~1, no attempt 
 has been made to apply a k-correction to the $B-R$ colours. The error-bars 
 shown in Fig.~3c (and in Fig.~4) take account of poisson statistics and 
 the estimated 15\% uncertainty in the background counts due to field-to-field variance.

\begin{figure}
  \epsfig{file=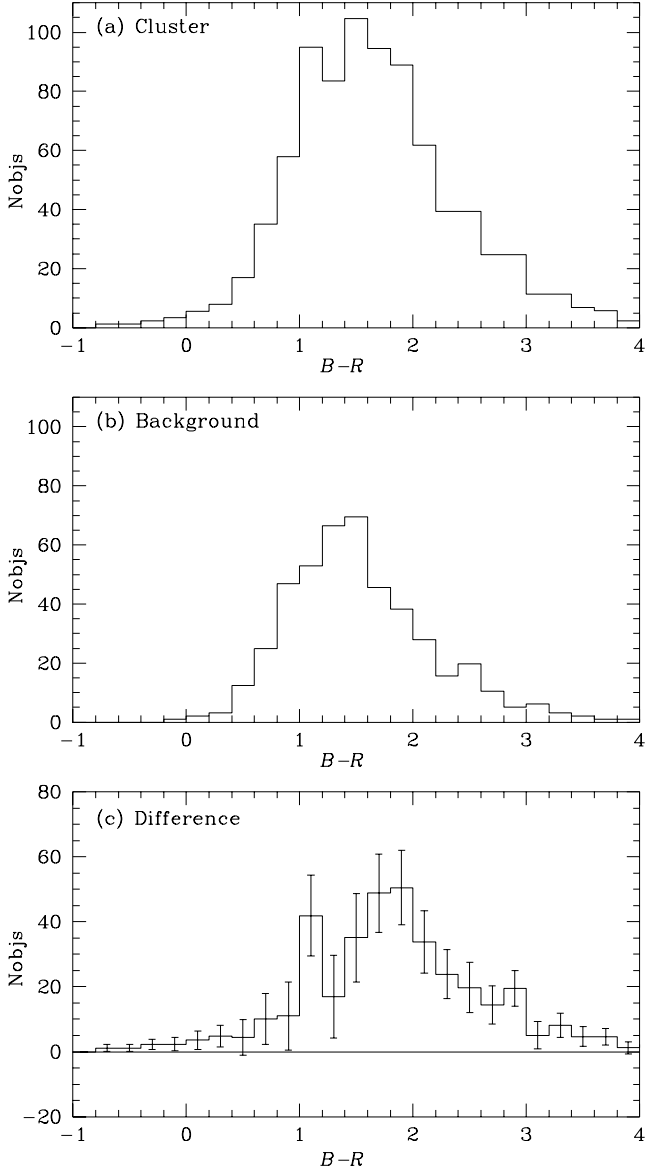}
  \caption{$B-R$ histograms for galaxies with  21.02$<R<$23.52 (equivalent to --18.0$<M_{R}<$--15.5 at cluster distance) in  (a) Cluster frame, (b) Background frame (c) Cluster - 
 Background.}
\end{figure}

The background-subtracted cluster $B-R$ colour distribution  
 has a sharp pronounced peak 
 at $B-R\simeq$1.1 and a much broader peak at $B-R\simeq$1.8.  
 Based upon the k-corrections of Coleman et al. (1980) and 
  Trentham (1998) we would expect dIrr galaxies at the distance of A868 
 to have observed $B-R\simeq$1.1 and dEs to cover a broader range 
 of observed colours 1.6$<B-R<$2.3. Clearly, 
  these values match well with the observed peaks 
 seen in Fig.~3c, suggesting that both dE and dIrr galaxies are 
    contributing significantly to the faint end of the LF of A868.

The relative contributions of dE and dIrr galaxies to the LF at $M_{R}>$--18.0 
 can be better seen in Fig.~4. This shows the background-subtracted  
  $B-R$ colour distribution for two absolute magnitude ranges:  
  --18.0$<M_{R}<$--16.5 
 and --16.5$<M_{R}<$--15.5.  In the brighter of these ranges the dEs clearly 
 numerically dominate the dIrrs. However, in the fainter range there are similar 
 numbers of dIrrs to dEs. The two populations can also be clearly seen on 
 the colour-magnitude 
diagram of Fig.~2: the dEs at 
  $R\sim$21.5, $B-R\sim$1.6 and the dIrrs  at  $R\sim$22.0, $B-R\sim$1.1.

\begin{figure}
  \epsfig{file=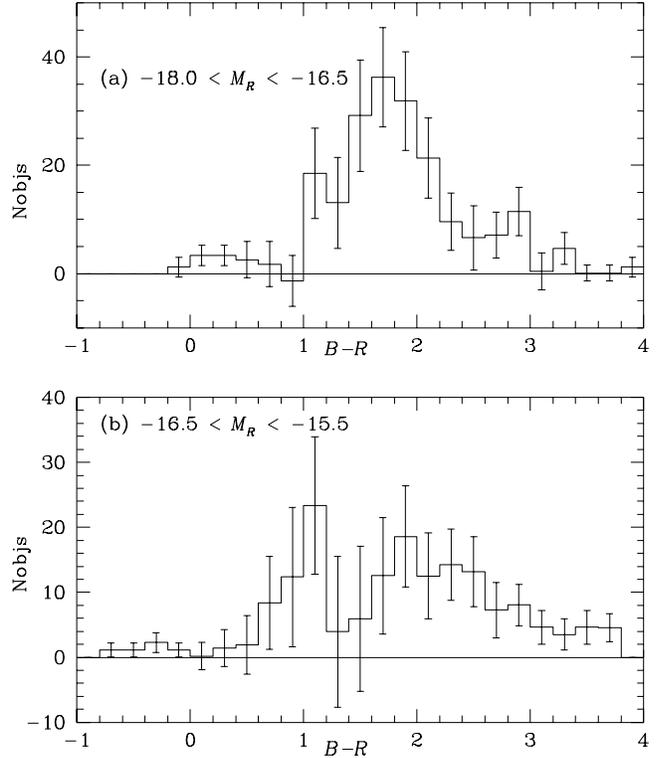}
  \caption{Background-subtracted $B-R$ colour distribution for A868:  
   (a) --18.0$<M_{R}<$--16.5; (b) --16.5$\le$$M_{R}<$--15.5.}
\end{figure}

The background-subtracted colour distributions of 
 Fig.~3 and Fig.~4 contain a significant number of galaxies which 
 are too red (i.e. $B-R>$2.5) to be plausibly considered cluster dEs, 
although the random errors in $B-R$ ($\simeq$0.3~mags) imply 
  that some galaxies 
 beyond $B-R$$>$2.5 will be cluster members. There is, however, the suggestion 
of a peak in the $B-R$ distribution at $B-R$$\simeq$2.9 which cannot be easily explained at being due to cluster members. A possible explanation for this is that it results from a cluster along the line of sight behind A868. 
  For example, a gE galaxy of $M_{R}$=--21  at $z$=0.45 would have 
 an apparent $R$ magnitude of around 21.2 and an observed $B-R\simeq$2.9 
  and would contribute to the 
 peak seen at $B-R$=2.9. The contribution of such a cluster 
 to the number counts would fall within the estimated 15\% uncertainty
 due to field-to-field variance. However, the gE sequence associated with
 such a cluster would show up as a significant peak 
 in the subtracted $B-R$ colour distribution, because these galaxies 
 inhabit a very narrow intrinisic range of $B-R$ colours. This would  
 not represent a general under-representation of the field-to-field 
 variance throughout the entire subtracted $B-R$ colour distribution. 
 The contribution of 
 background structure to the $B-R$ distribution cannot be significant 
 at $B-R<$2.5. As noted above, the cluster ellipticals have a 
 sharply peaked distribution at $B-R\simeq$2.2. Any bright background 
 galaxies would have to lie at redder colours than this and be sufficiently 
  distant to lie at $R>$21.02. 

\section{Discussion and conclusions}

In this section we consider the relative contributions 
 to the faint-end of the cluster LF of the three populations of 
  galaxies identified in Section~3. To study this 
 we split the background-subtracted $B-R$ colour distribution 
 shown in Fig.~3c into three groups. All galaxies with  
 $B-R<$1.4 are assumed to be dIrrs. All 
 galaxies with 1.4$\le$$B-R$$\le$2.5 are assumed to be dEs. 
 All galaxies with $B-R>$2.5 are assumed to be background gEs.

Fig.~5a reproduces the 
  LF of Fig.~1 (solid squares). Also shown (open squares) is a LF derived by excluding (for 
  galaxies with $R>$21.02) all the background gEs. 
    Excluding these
 objects does not significantly alter 
  the slope of the faint-end upturn in the LF
  to $\alpha$=--1.22$\pm$0.16. Hence, although there is evidence that 
 background large-scale structure is contributing to the counts at the faint-end of the 
 derived LF for A868, this contribution is nowhere near sufficient to explain, 
 by itself, the steep upturn seen at the faint-end of the cluster LF.

\begin{figure}
  \epsfig{file=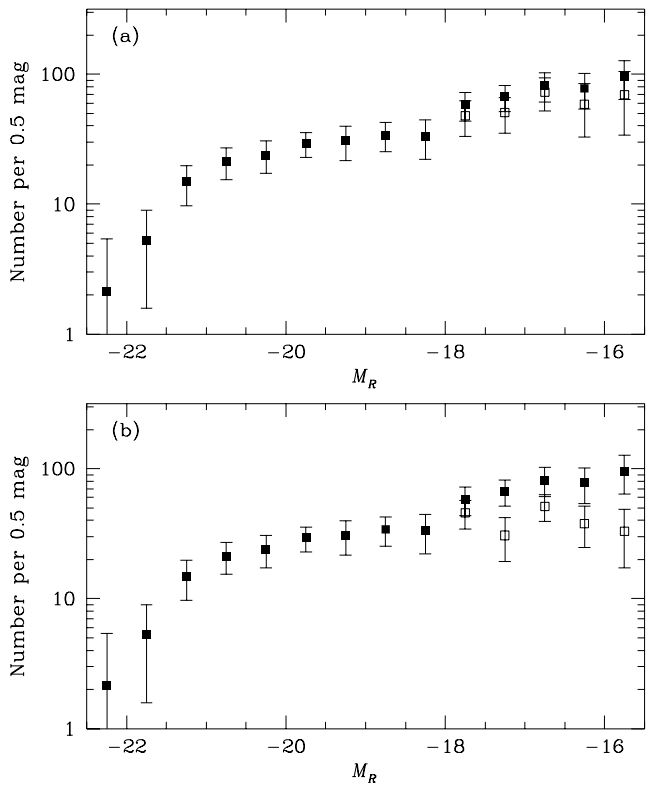}
  \caption{(a) The solid squares show the R-band LF plotted in Fig.~1. 
 The open squares show the LF derived if  objects with $R>$21.02 (i.e. $M_{R}>$--18.0 if 
 they are in the cluster)  
 and $B-R>$2.5 (those identified as background gEs) are excluded; (b) The solid-squares 
 again show the R-band LF plotted in Fig.~1. The open squares show the LF derived if 
 objects with $R>$21.02 and $B-R<$1.4 or $B-R>$2.5 (those identifed as cluster 
 dIrrs or background gEs) are excluded. }
\end{figure}

In Fig.~5b, we again reproduce the LF of Fig. ~1 (solid squares). However, we now show
 (open squares) the LF derived by excluding not only the red background gEs but also 
 the dIrr galaxies. Whilst there is significant scatter in these points, 
  it is clear that there is now no evidence for an 
 upturn in the LF of A868 at faint magnitudes ($\alpha$=--0.91$\pm$0.16). 
 The upturn in the cluster LF depends on 
 the presence of both the dE and dIrr populations.

\begin{figure}
  \epsfig{file=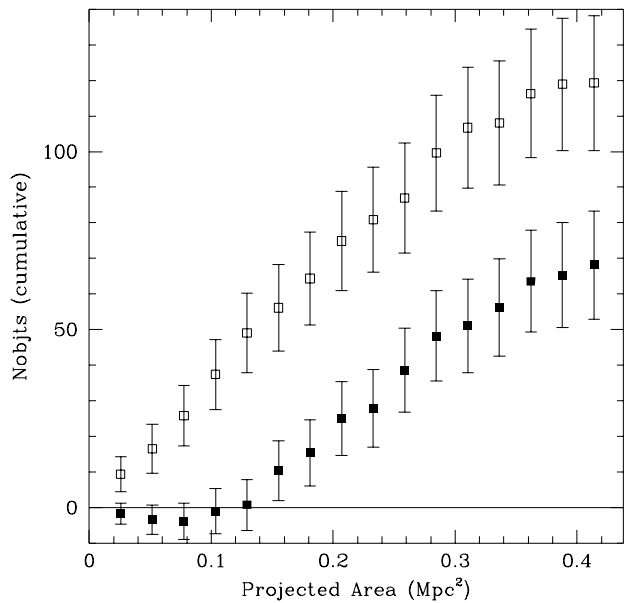}
  \caption{The open squares show the excess number of dE galaxies (defined as having 
 $R>$21.02, 1.4$\le$$B-R$$\le$2.5) over background objects as a cumulative function 
 of cluster projected area. The solid squares show the excess number of dIrr galaxies (defined as having 
 $R>$21.02, $B-R<$1.4) over background objects as a cumulative function 
 of cluster projected area.} 
\end{figure}

Figure~6 shows the number  of dE and dIrr objects as a cumulative 
 function of projected area from the cluster centre (taken to be the 
 centre of the cD galaxy). These numbers were derived by 
 calculating the number of objects within each 
  $B-R$ range in a series of annuli which linearly 
 increase in projected area. The counts for each annulus were 
 adjusted by subtracting from them the average number of background 
  counts expected in an annulus of that projected area (calculated 
  from the background frame). 

Perhaps the most striking feature of Fig.~6 is the apparent 
 absence of any dIrr galaxies inside a projected area of $\simeq$0.12~Mpc$^{2}$
 (equivalent to a projected radius of $\simeq$0.19~Mpc). In contrast there is a significant 
 excess of dE galaxies 
 over background counts right into the core of the cluster. 
  This result is consistent with the `galaxy harassment' scenario 
 (Moore et al. 1996, 1998). These simulations show 
 that the morphology of brighter disk systems in a dense cluster 
 can be radically transformed 
 by the effects of  close encounters with brighter galaxies and the 
 cluster's tidal field. Gas and stars are  progressively stripped  
    out  of the disk systems eventually
 leaving a spheroidal remnant (Moore et al. 1998). The processes 
of `harassment' would certainly fade dIrr galaxies by stripping 
 gas and stars from them and may eventually leave a  faint 
 dE remnant.  
 In such `harassment' scenarios
  the material at the cluster centre is older than material in the outer regions. 
 Hence objects in the cluster centre experience `harassment' at earlier times 
 and for longer durations. This could explain the absence of dIrrs
 in the core of A868.

 The second interesting feature of Fig.~6 is that the ratio of dE 
 galaxies to dIrrs appears similar at projected radii $>$0.19~Mpc. This suggests 
that the population of dIrrs
 cannot be explained as being due 
 to  recent infall. The populations of dE and dIrr 
 would have to have formed part of the cluster for  similar 
 timescales in order to achieve similar radial distributions. 
  The dEs numerically dominate down to our completeness
 limits. However, as can be seen from Fig.~4b, it is possible that the 
 dIrrs will dominate at fainter magnitudes. 
   
The third point  
 from Fig.~6 is that there appears to be a down-turn in  
 the number densities of both dE and dIrr galaxies beyond a projected area 
 of $\simeq$0.38~Mpc$^{2}$ (projected radii $>$0.35~Mpc), 
 although the errors are so large at these radii that any such 
 conclusion has to be tentative. 
 Are we seeing the edge of the cluster's dwarf halo ? At what point 
 do the number densities fall to field levels ? 
  Spectroscopic surveys  are the best way to study 
   cluster LFs at large cluster radii where the decreasing number of 
 cluster galaxies makes statistical subtraction studies problematical.

 Phillipps et al. (1998) noted that the steepness of the 
  faint-end slope of the LF appears to be dependent on cluster 
 density, with the ratio of 
 dwarfs-to-giants increasing in lower density environments.
 Driver et al. (1998b) noted a tendency for less evolved clusters   
 (including A868) to have steeper faint-end slopes compared to 
 more evolved clusters. 
   A possible scenario is that 
 clusters form with large numbers of both dE and dIrr galaxies and that
  such unevolved  clusters will have LFs with
   steep faint-end slopes. The various 
  potential dynamical processes in the cluster (e.g. harassment, ram-pressure stripping 
 etc.) preferentially strip gas and stars from the less centrally concentrated 
  dIrr galaxies leading to their gradual 
 fading and with it a lessening of the faint-end slope of the LF. This happens firstly in the 
 cluster core where such processes have had 
 a longer period to operate but ultimately, 
 in very evolved systems, the dIrrs throughout the cluster 
 are `faded' and a flat faint-end slope to the LF at all radii results.

\section*{Acknowledgments}

PJB and JBJ acknowledge the financial support of the UK PPARC. 
 WJC acknowledges the financial support of the Australian Research Council
 during the course of this work.  
This research has made use of the NASA/IPAC Extragalactic Database (NED) which is
operated by the Jet Propulsion Laboratory, Caltech, under agreement with the 
 National Aeronautics and Space Administration.

\end{document}